\documentclass{svmult}

\usepackage{ae,aecompl}
\usepackage{makeidx}      
\usepackage{graphics}
\usepackage{graphicx}
\usepackage{multicol}
\usepackage{enumerate}
\usepackage[bottom]{footmisc}
\usepackage{pifont}
\usepackage{amsmath}
\usepackage{amssymb}
\usepackage[sort&compress]{natbib}

\makeindex 

\def\ev #1{\left\langle #1 \right\rangle}

\begin{document}

\title*{Why do Hurst exponents of traded value increase as the logarithm of company size?}
\titlerunning{Why do Hurst exponents of traded value increase as the logarithm of\dots}
\author{Zolt\'an Eisler\inst{1} \and J\'anos Kert\'esz \inst{1,2}}

\institute{Department of Theoretical Physics, Budapest University of Technology and Economics, Budafoki \'ut 8, H-1111 Budapest, Hungary \and Laboratory of Computational Engineering, Helsinki University of Technology, P.O.Box 9203, FIN-02015 HUT, Finland}

\maketitle

\begin{abstract}
The common assumption of universal behavior in stock market data can sometimes lead to false conclusions. In statistical physics, the Hurst exponents characterizing long-range correlations are often closely related to universal exponents. We show, that in the case of time series of the traded value, these Hurst exponents increase logarithmically with company size, and thus are non-universal. Moreover, the average transaction size shows scaling with the mean transaction frequency for large enough companies. We present a phenomenological scaling framework that properly accounts for such dependencies.
\end{abstract}

\keywords{econophysics; stock market; fluctuation phenomena}\vskip20pt

\section{Introduction}

The last decades have seen a lot of contribution by physicists to various other subjects. The applied methods are often rooted in modern statistical physics, and in particular scaling and universality. Examples range from biology \cite{vicsek.biol} to finance \cite{evolving, kertesz.econophysics, bouchaud.book, stanley.book}. Successes achieved in the latter area have lead to the formation of a whole new field, commonly called econophysics. But despite the large number of studies and the undeniable progress that has been made, one worry still remains: There is very little theoretical ground to assume that physical concepts are actually appropriate to describe, e.g., stock market fluctuations \cite{gallegatti.etal}. Critically speaking, it is equally justified to consider the often used power laws as only a way to fit data. Instead of universality, what one actually observes in econophysics, can also be seen as just a robustness of qualitative features, which is a much weaker property.

In this paper we revisit a previously introduced framework for financial fluctuations \cite{eisler.unified}, that can be used to explicitly show the absence of universal behavior in trading activity. The paper is organized as follows. Section $2$ introduces notations and the dataset that will be used. Section $3$ shows, that many key features of the traded value of stocks depend on the size of the company whose stock is considered. We find that as we go to larger and larger companies:
\begin{enumerate}[(i)]
\item the average transaction size increases,
\item the Hurst exponent of traded value/min grows as the logarithm of the mean of the same quantity,
\item fluctuations of the trading activity grow as a non-trivial, time scale dependent power of mean traded value.
\end{enumerate}
Section $4$ integrates these findings into a consistent, common scaling framework, and points out the connection between the three observations. 

\section{Notations and data}
\label{sec:notations}
First, let us introduce a few notations that will be used throughout the paper. For a time window size $\Delta t$, one can write the total traded value of stock $i$ during the interval $[t, t+\Delta t]$ as
\begin{equation}
f_i^{\Delta t}(t) = \sum_{n, t_i(n)\in [t, t+\Delta t]} V_i(n),
\label{eq:flow}
\end{equation}
where $t_i(n)$ is the time of the $n$-th transaction of stock $i$. The number of elements in the sums, i.e., the number of trades in the time window, we will denote as $N_{\Delta t}(t)$. The 
so called tick-by-tick data are denoted by $V_i(n)$, which is the value exchanged in
trade $n$. This is the product of the transaction price $p_i(n)$ and the
traded volume $\tilde V_i(n)$:
\begin{equation}
V_i(n) = p_i(n) \tilde V_i(n).
\label{eq:v}
\end{equation}
Note that the use of $V$ instead of $\tilde V$ automatically eliminates any anomalies caused by stock splits or dividends.

The data we analyze is from a TAQ database \cite{taq1993-2003}, containing all transactions of the New York Stock Exchange (NYSE) for the years $1993-2003$. The samples were always restricted to those stocks that were traded every month during the period of that specific calculation. We detrended the data by the well-known $U$-shaped daily pattern of traded volumes, similarly to Ref. \cite{eisler.non-universality}.

Finally, $\ev{\cdot}$ always denotes time average, and $\log(\cdot)$ means $10$-base logarithm throughout the paper.

\section{Size dependent properties of trading activity}
\label{sec:sizedep}

In the econophysics literature, it is common practice to assume a form of universal behavior in stock market dynamics. The trading of different stocks, on different markets and for various time periods is assumed to follow the same laws, and this is -- at least qualitatively -- indeed found in the case of many stylized facts \cite{stanley.book, bouchaud.book}. However, recent studies \cite{eisler.sizematters, eisler.unified} have pointed out, that this is not completely general. In this section, we present two properties of trading, that appear robust between markets and time periods, and which are related to a distinct company size dependence.

Company size is usually measured by the capitalization, but trading frequency $\ev{N_{\Delta t}}$ (measured in trades/min), or the average traded value $\ev{f_{\Delta t}}$ (measured in USD/min) are also adequate measures of the importance of a company: Very small companies are traded infrequently, while large ones very often, and, naturally, traded value has a corresponding behavior. In fact, one finds, that $\ev{N_{\Delta t}}$ and $\ev{f_{\Delta t}}$ are non-trivial, monotonic functions of capitalization \cite{eisler.sizematters, zumbach}.

\subsection{Dependence of the average trade size on trading frequency}
\label{sec:stick}
\begin{figure}[tbp]
\centerline{\includegraphics[height=180pt]{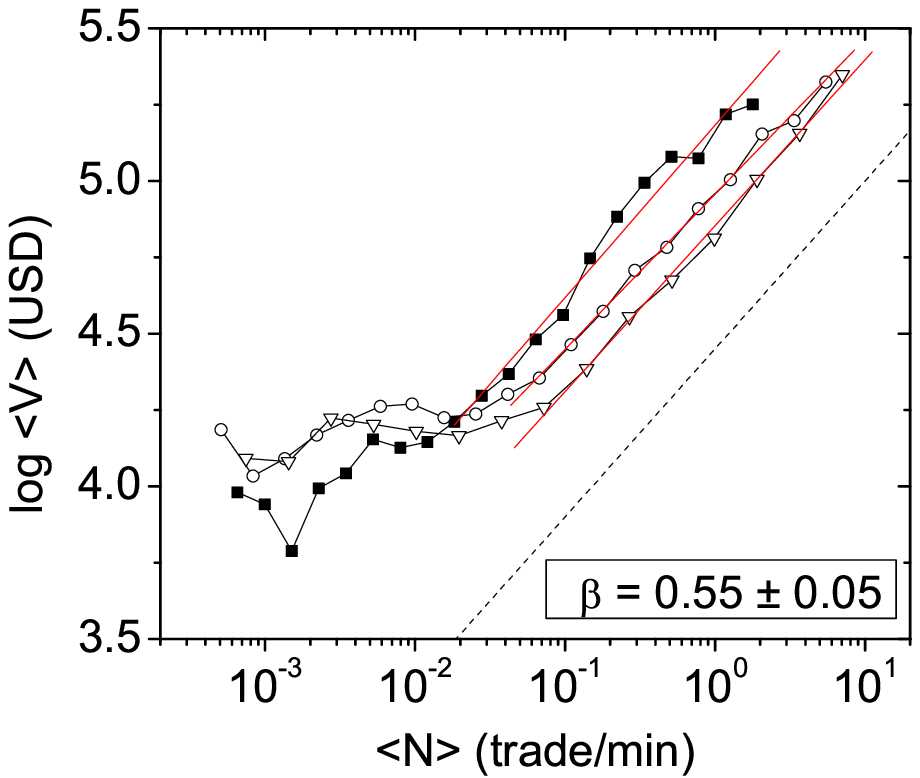}}
\caption{The dependence of the mean value per trade $\ev{V_i}$ on the
average number of trades/min $\ev{N_i}$. Calculations were done for the
periods $1994-1995$ ($\Box$), $1998-1999$ (O), and for the year
$2000$ ($\bigtriangledown$, see Ref. \cite{eisler.sizematters}). For the smallest stocks the data lack a clear tendency. However, larger stocks show scaling between the two quantities, according to \eqref{eq:vvsn}. The slopes are around $\beta = 0.55 \pm 0.05$, regardless of time period. \emph{Note}:
Groups of stocks were binned, and $\log \ev{V_i}$ was averaged for better
visibility.}
\label{fig:mNvsmpV}
\end{figure}

Let us first construct a very simple measurement: calculate the average number of trades per minute ($\ev{N_i}$) and the mean value exchanged per trade ($\ev{V_i}$). One can plot these two quantities versus each other for all stocks (see Fig. \ref{fig:mNvsmpV}), to find a remarkably robust behavior. For all the periods $1994-1995$, $1998-1999$, and $2000$, the data lack a clear tendency where trading frequency is low ($\ev{N_i} < 10^{-2}$ trades/min). Then, as we go to more frequently traded companies, an approximate power law emerges:
\begin{equation}	
\ev{V_i} \propto \ev{N_i}^\beta .
\label{eq:vvsn}
\end{equation}
The measured exponents are around $\beta \approx 0.5$, systematically greater than the value $\beta \approx 0.2$ found for NASDAQ (see also Refs. \cite{eisler.sizematters, eisler.unified}), and smaller than $\beta \approx 1$ for London's FTSE-100 \cite{zumbach}.

In some sense trades appear to "stick together": Once a stock is traded more and more intensively, traders seem to prefer to increase their size as the frequency cannot be increased beyond limits.

\subsection{Size-dependent correlations}
\label{sec:hurst}
The correlation properties of stock market time series have been studied extensively \cite{gopi.volume, bouchaud.book, stanley.book, cont.stylized}. However, with very few exceptions \cite{bonanno.dynsec}, such studies were limited to the stocks of large companies. Those, in general, were found to display universal patterns of behavior.

In this section we focus on the correlations of the traded value $f$. Recently it was pointed out by two independent studies \cite{queiros.volume, eisler.sizematters} that both this $f$ and trading volumes have finite variance, in contrast to early findings \cite{gopi.volume}. Thus, it is meaningful to define a Hurst exponent $H(i)$ \cite{vicsek.book, dfa} for $f$ as
\begin{equation}
\sigma(i, \Delta t) = \ev{\left ( f_i^{\Delta t}(t) - \ev{f_i^{\Delta t}(t)}.
\right )^2} \propto \Delta t^{H(i)}, \label{eq:hurst}
\end{equation}
The signal is correlated for $H>0.5$, and uncorrelated for $H=0.5$. Significant anticorrelated behavior ($H<0.5$) does
not usually occur in this context.

One finds, that the Hurst exponent does not exist in a strict sense: all
stocks show a crossover \cite{eisler.sizematters}
between two types of behavior around the time scale of $1$ day. This
threshold depends on the market and the time period under study, but keeping those constant,
it does not depend on the actual stock in any systematic way.

\begin{figure}[tbp]
\centerline{\includegraphics[height=190pt]{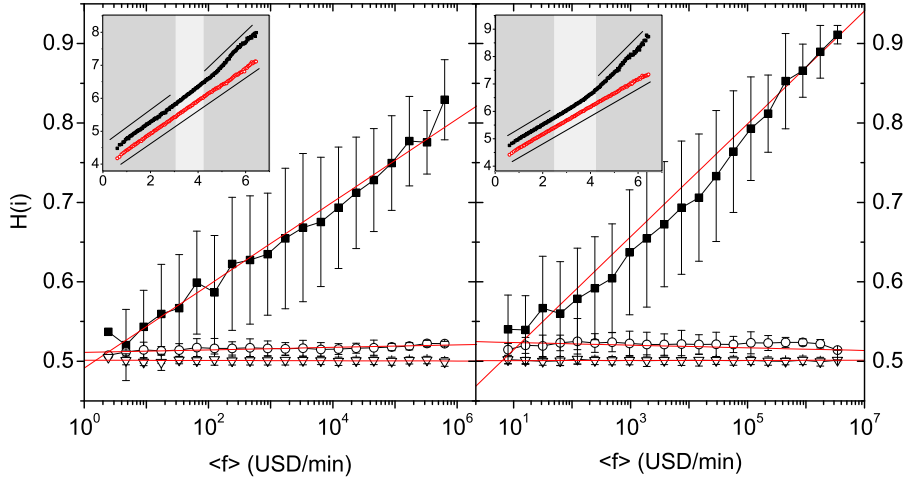}}
\caption{Value of the Hurst exponents $H(i)$ for the time
periods $1994-1995$ (left) and $2000-2002$ (right). For short time windows (O, $\Delta t < 20$ min for $1994-1995$, and $\Delta t < 6$ min for $1998-1999$), all signals are nearly uncorrelated, $H(i)\approx 0.51 - 0.52$. The fitted slope is $\gamma_{t}=0.00\pm 0.01$. For larger time windows ($\blacksquare$. $\Delta t > 300$ min), the strength of correlations depends logarithmically on the mean trading activity of the stock, 
$\gamma_{t}=0.05\pm 0.01$ for $1994-1995$, and $\gamma_{t}=0.05\pm 0.01$ for $1998-1999$. Shuffled data ($\bigtriangledown$) display no correlations, thus $H_{\mathrm{shuff}}(i) = 0.5$, which also implies $\gamma_{t} = 0$. {\it Note}: Groups of stocks were binned, and their logarithm was averaged. The error bars show standard deviations in the bins. {\it Insets:} The
$\log \sigma$-$\log \Delta t$ scaling plots for Wal-Mart (WMT, $\blacksquare$). The
darker shaded
intervals have well-defined Hurst exponents, the crossover is indicated with a
lighter background.
The slopes corresponding to Hurst exponents are $0.52$ and $0.73$ for $1994-1995$, and $0.52$ and $0.89$ for $1998-1999$. The slope for shuffled data is $0.5$. Shuffled points (O) were shifted vertically for better visibility. }
\label{fig:hurst}
\end{figure}

We did the calculations for two time periods, $1994-1995$ and $1998-1999$. 
Under a certain size of time windows, which is $\Delta t < 20$ min for $1994-1995$ and
$\Delta t < 6$ min for $1998-1999$, the trading activity is uncorrelated for 
all stocks. However, when one chooses $\Delta t > 300$ min, the picture changes completely.
There, small $\ev{f}$ stocks again display only very weak correlations, but larger ones up to $H\approx
0.9$. Moreover, there is a clear logarithmic trend in the data:
\begin{equation}
H(i) = H^* + \gamma_{t}\log\ev{f_i},
\end{equation}
with $\gamma_{t}(\Delta t > 300$ min$) = 0.05\pm0.01$ for $1994-1995$ and
$\gamma_{t}(\Delta t > 300$ min$) = 0.07\pm0.01$ for $1998-1999$. As a reference, we also checked that $H_{\mathrm{shuff}}(i)=0.5$ for the shuffled time series. All results are shown in Fig. \ref{fig:hurst}.

The most interesting point is that the crossover is not from \emph{uncorrelated} to \emph{correlated}, but from \emph{homogeneous} to \emph{inhomogeneous} behavior. For short times, all stocks show $H(i)\approx H_1$, i.e., $\gamma_{t} = 0$. For long times, $H(i)$ changes with $\ev{f_i}$ and $\gamma_{t} > 0$. This can also be understood as a dependence on company size, as $\ev{f}$ is roughly proportional to capitalization \cite{eisler.sizematters}.

\subsection{Fluctuation scaling}

The technique of fluctuation scaling is very similar to the above, and it was recently applied to stock market data (see, e.g., Refs. \cite{eisler.non-universality, eisler.unified}). It is based on a phenomenological scaling law that connects the standard deviation $\sigma_i$ and the average $\ev{f_i}$ of the trading activity for all stocks:
\begin{equation}
\sigma(i, \Delta t) \propto \ev{f_i}^{\alpha (\Delta t)},
\label{eq:alpha}
\end{equation}
where the scaling variable is $\ev{f_i}$ (or $i$), and $\Delta t$ is kept constant. That is, the standard deviation of a quantity scales with the mean of the same quantity. $\sigma(i, \Delta t)$ is the same as used in the definition of the Hurst exponent \eqref{eq:hurst}, where $i$ was constant and $\Delta t$ was varied. 

\begin{figure}[ptb]
\centerline{\includegraphics[height=160pt]{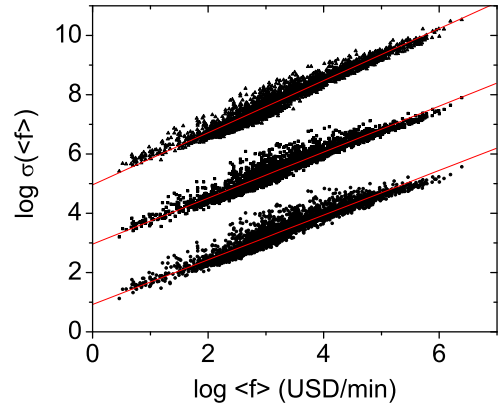}}
\caption{Examples of $\sigma(\ev{f})$ scaling plots for NYSE, period
$1998-1999$. The window sizes from bottom to top: $\Delta t = 10$ sec,
$0.5$ day, $2$ weeks. The slopes are $\alpha(\Delta t) = 0.75, 0.78, 0.88$,
respectively. Points were shifted vertically for better visibility.}
\label{fig:alphaexampleNYSE}
\end{figure}

The presence of scaling \eqref{eq:alpha} is not at all a trivial fact. Nevertheless, one finds that it holds quite generally, for any $\Delta t$. Here, we confirm this for the periods $1994-1995$ and $1998-1999$, examples of scaling plots are shown in Fig. \ref{fig:alphaexampleNYSE}.

\begin{figure*}[ptb]
\centerline{\includegraphics[height=175pt]{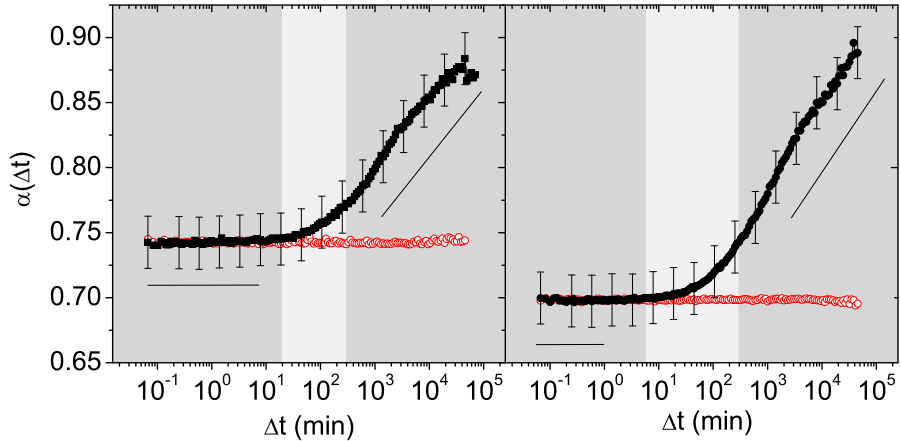}}
\caption{The dependence of the scaling exponent $\alpha$ on the window
size $\Delta t$. The darker shaded intervals have well-defined Hurst exponents
and values of $\gamma_{t}$, the crossover is indicated with a lighter background.
$1994-1995$ (left): without shuffling ($\blacksquare$)
the slopes of the linear regimes are $\gamma_{f}(\Delta t<20$
min$)=0.00\pm 0.01$ and $\gamma_{f}(\Delta t>300$ min$)=0.05\pm 0.01$. For
shuffled data (O) the exponent is independent of window size,
$\alpha (\Delta t)=0.74\pm0.02$. $1998-1999$ (right): without shuffling
($\blacksquare$) the slopes of the linear regimes are $\gamma_{f}(\Delta
t<6$ min$)=0.00\pm0.01$ and $\gamma_{f}(\Delta t > 300$
min$)=0.07\pm0.01$. For shuffled data (O) the exponent is
independent of window size, $\alpha (\Delta t)=0.70\pm0.02$.}
\label{fig:alpha}
\end{figure*}

A systematic investigation finds, that $\alpha$ strongly depends on the $\Delta t$ size of the time windows. Fig. \ref{fig:alpha} shows, that when $\Delta t$ is at most a few minutes, $\alpha(\Delta t)$ is constant, the values are $0.74$ and $0.70$ for $1994-1995$ and $1998-1999$, respectively. Then, after an intermediate regime, for window sizes $\Delta t > 300$ min, there is a logarithmic trend:
\begin{equation}
\alpha(\Delta t) = \alpha^* + \gamma_{f}\log \Delta t,
\end{equation}
with slopes $\gamma_f = 0.05\pm 0.01$ for $1994-1995$, and $\gamma_f = 0.07\pm 0.01$ for $1998-1999$.

\section{Scaling theory}
\label{sec:scaling}

The stock market is only one of many examples for fluctuation scaling, which is in fact a very general and robust phenomenon, that was observed in many complex systems, ranging from highway traffic to fluctations in the visitations of web pages \cite{barabasi.fluct, eisler.non-universality, eisler.internal}. Here, the elements of the system are the web pages or highways, and $f_i(t)$ is not trading activity, but still some a form "activity" (number of visitation to the page, volume of car traffic through the road). 

A previous study \cite{barabasi.fluct} found, that in temporally uncorrelated systems, two universality classes exist. Under strong external driving, all systems show $\alpha = 1$. Systems with a robust internal dynamics, consisting of i.i.d. events, display $\alpha = 1/2$.

When the size of the events is not identically distributed throughout the system, that can lead to the breaking of universality, and intermediate values of $\alpha$. This is what happens in the case of stock markets. When $\Delta t$ is small (seconds to a few minutes), the transactions can be assumed to arrive independently, but their size is inhomogeneous, as pointed out in Sec. \ref{sec:stick}. In the complete absence of correlations, there would a clear relationship between the exponents $\alpha$ and $\beta$ \cite{eisler.internal}:
\begin{equation}
\alpha = \frac{1}{2}\left(1+\frac{\beta}{\beta+1}\right ).
\label{eq:internal}
\end{equation}

Substituting $\beta = 0.55$ yields $\alpha(\Delta t \rightarrow 0) = 0.68 \pm 0.01$, which agrees within the error bars with the result for $1998-1999$, but it is somewhat smaller than the actual value for $1994-1995$. Also note that $\beta$ only exists for large enough stocks, whereas $\alpha$ and fluctuation scaling applies to all stocks. We believe, that the discrepancies are due to the fact, that the picture presented in Ref. \cite{eisler.internal} is an overly simplified model for the stock market. Nevertheless, it is remarkable, that the breaking of universality and the rough value of the exponent is predicted properly.

Let us now turn to the $\Delta t$ dependence of $\alpha$. First of all, let us notice, that for both periods, there is a change from a constant value to a logarithmic increase, and this is at exactly the same $\Delta t$'s, where Sec. \ref{sec:hurst} found the crossovers from homogeneous to inhomogeneous correlations. In fact, the correspondence between the two observations is not incidental. Both the Hurst exponents $H(i)$ and $\alpha(\Delta t)$ describe the behavior of the same standard deviation $\sigma(i, \Delta t)$: $$\sigma(i, \Delta t) = \ev{\left ( f_i^{\Delta t}(t) - \ev{f_i^{\Delta t}(t)} \right )^2} \propto \Delta t^{H(i)},$$ and $$\sigma(i, \Delta t) \propto \ev{f_i}^{\alpha (\Delta t)}.$$ A simple calculation \cite{eisler.unified} can show that the only possible way for these two scaling laws to coexist, is when
\begin{eqnarray}
\alpha(\Delta t) = \alpha^*+\gamma\log \Delta t\\
H(i) = H^*+\gamma\log \ev{f_i},
\end{eqnarray} 
where a key point is that the two slopes $\gamma$ are the same. In short, for the two previously introduced constants $\gamma_t = \gamma_f = \gamma$.

Again, this is in harmony with the actual observations. Due to the previously mentioned crossover in correlations, one has to distinguish three regimes in $\Delta t$.
\begin{enumerate}
\item For small $\Delta t$, all stocks display the same, nearly uncorrelated trading behavior, i.e., $\gamma = 0$. Accordingly, $\alpha(\Delta t)$ is constant, regardless of window size.
\item For an intermediate range of $\Delta t$'s, we are in the crossover regime. $H$ does not exist for any stock. $\alpha$ still does exist, but -- as expected -- its time window dependence does not follow a logarithmic trend.
\item For large $\Delta t$, the Hurst exponent increases logarithmically with the mean traded value $\ev{f}$, and so does $\alpha$ with $\Delta t$. The slopes agree very well ($\gamma_t = \gamma_f$) for both time periods.
\end{enumerate}

As noted before, the equality $\gamma_t = \gamma_f$ can be calculated fairly easily, but one can look at this result in a different way. Both fluctuation scaling and the Hurst exponent (or equivalently, power law autocorrelations) are present in a very wide range of complex systems. But we have just seen that this is only possible in two of ways: the correlations must either be homogeneous throughout the system ($H(i)=H$, $\gamma = 0$), or they must have a logarithmic dependence on mean activity. Consequently, when one for example looks at the results of Sec. \ref{sec:hurst}, they are not surprising at all. The coexistence of our two scaling laws is so restrictive, that if the strength of correlations depends on company size, and thus on $\ev{f}$, the realized logarithmic dependence is the only possible scenario.

\section{Conclusions}

In the above, we presented some recent findings concerning the fluctuations of stock market trading activity. As the central point of the paper, we discussed the application of fluctuation scaling. We gave evidence, that the logarithmic increase of the Hurst exponent of traded value with the mean traded value comes as a very natural consequence of fluctuation scaling. The behavior of the companies depends on a continuous parameter: the average traded value, or equivalently, company size.

This is a clear sign of non-universality, and thus contrary to a naive expectation from statistical physics. For example, in the case of surface growth \cite{vicsek.book}, the Hurst exponent of surface height fluctuations is universal to a certain type of growth dynamics. In contrast, on the market the "dynamics" (i.e., trading rules) are the same for all stocks, but the resulting exponents still vary. While we believe that it is possible, that at least some properties of trading are universal, but we wish to point out that not \emph{all} of them are.

Our results imply that one must take great care when applying concepts like
scaling and universality to financial markets. The present theoretical models of trading should be
extended to account for the capitalization dependence of the characteristic quantities, which is
a great challenge for future research.

\section*{Acknowledgments}

The authors would like to express their gratitude to Bikas K. Chakrabarti, Arnab Chatterjee and all organizers of the International Workshop on the Econophysics of Stock Markets and Minority Games for their infinite hospitality. They are also indebted to Gy\"orgy Andor for his support with financial data. JK is member of the Center for Applied Mathematics and Computational
Physics, BME. Support by OTKA T049238 is acknowledged.

\bibliographystyle{unsrt}
\bibliography{kolkataproc3}

\end{document}